\documentclass{article}

\usepackage{spconf}
\usepackage[T1]{fontenc}
\usepackage{amsmath, amssymb}
\usepackage{newtxtext, newtxmath}
\usepackage[utf8]{inputenc}

\newcommand{\bm}[1]{\mathbf{#1}}

\DeclareFontFamily{T1}{qcr}{\hyphenchar\font-1}
\DeclareFontShape{T1}{qcr}{m}{n}{<-> s*[0.85]ec-qcrr}{}
\usepackage[english]{babel}
\usepackage{microtype}
\usepackage{expl3, xparse}
\usepackage{graphicx}
\usepackage{subfig}
\usepackage{setspace}
\usepackage[hidelinks]{hyperref}
\usepackage{etoolbox}
\usepackage{layout}
\usepackage{smart-eqn}
\usepackage{array}
\usepackage{caption}
\usepackage{hhline}
\usepackage[absolute]{textpos}
\usepackage{tikz}
\usepackage{makecell}
\usepackage{placeins}
\usepackage{booktabs}
\usepackage{nicefrac}
\usepackage[doi=false, backend=bibtex,style=ieee, maxcitenames=1, mincitenames=1, maxbibnames=10]{biblatex}
\usepackage{csquotes}
\usepackage{xcolor}
\usepackage[acronym]{glossaries}
\usepackage{scrextend}
\usepackage[capitalise, noabbrev]{cleveref}

\usetikzlibrary{shapes}

\ExplSyntaxOn

\bgroup
\catcode`\(=1
\catcode`\)=2
\catcode`\{=12
\catcode`\}=12
\gdef\c_left_brace_str({)
\gdef\c_right_brace_str(})
\egroup

\bgroup
\catcode`\|=0
\catcode`\\=12
\gdef|c_backslash_str{\}
|egroup

\ExplSyntaxOff

\hypersetup{
pdftitle={Forensic Analysis and Localization of Multiply Compressed MP3 Audio Using Transformers},
pdfauthor={Ziyue Xiang, Paolo Bestagini, Stefano Tubaro, Edward J. Delp},
pdfkeywords={MP3 compression, audio forensics, convolutional neural networks, transformer networks},
pdfsubject={Machine learning, multimedia forensics, MP3 compression}
}

\addbibresource{refs.bib}

\newcommand{\symshield}{\includegraphics[height=1.2ex]{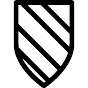}}
\newcommand{\symsword}{\includegraphics[height=1.6ex]{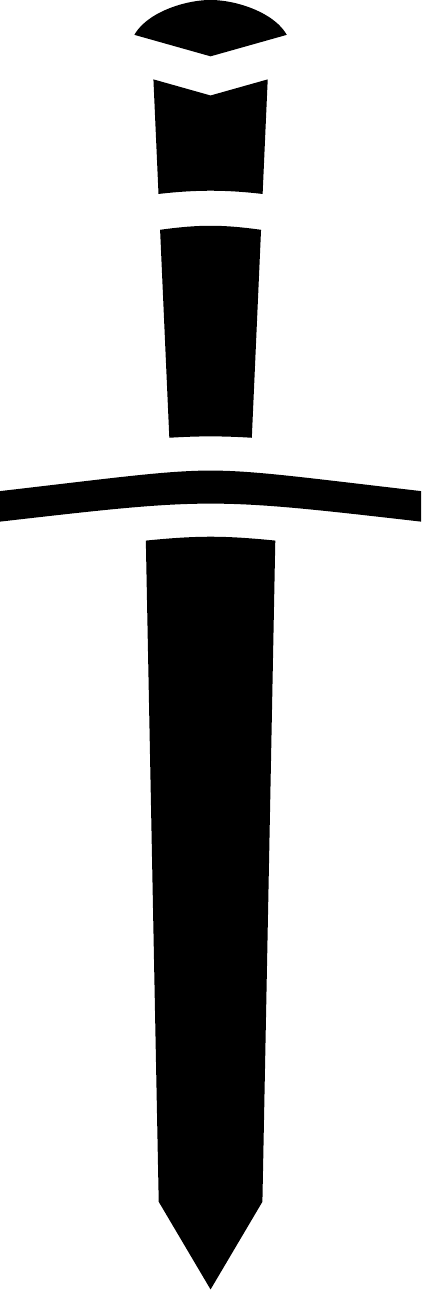}}

\newcolumntype{C}{>{\arraybackslash}c<{}}

\setkeys{glslink}{hyper=false}

\newacronym{pcm}{PCM}{Pulse Code Modulation}
\newacronym{fft}{FFT}{Fast Fourier Transform}
\newacronym{dnn}{DNN}{Deep Neural Network}
\newacronym{cnn}{CNN}{Convolutional Neural Network}
\newacronym{hhs}{HHS}{Human Hearing System}
\newacronym{mdct}{MDCT}{Modified Discrete Cosine Transform}
\newacronym{mlp}{MLP}{Multilayer Perceptron}
\newacronym{cbr}{CBR}{Constant  Bit Rate}
\newacronym{vbr}{VBR}{Variable Bit Rate}
\newacronym{aac}{AAC}{Advanced Audio Codec}
\newacronym{sa}{SA}{Self Attention}
\newacronym{msa}{MSA}{Multihead Self Attention}

\def\x{\bm{x}}
\def\y{\bm{y}}
\def\z{\bm{z}}
\def\Z{\bm{Z}}
\def\c{\bm{c}}
\def\p{\bm{p}}
\def\ifunc{\mathcal{I}}
\def\tfop{\operatorname{tf}}
\def\yhat{\hat{\y}}

\title{Forensic Analysis and Localization of \\Multiply Compressed MP3 Audio Using Transformers}

\name{Ziyue Xiang\textsuperscript{\symsword}, Paolo Bestagini\textsuperscript{\symshield}, Stefano Tubaro\textsuperscript{\symshield}, Edward J. Delp\textsuperscript{\symsword}\thanks{This material is based on research sponsored by the Defense Advanced Research Projects Agency (DARPA) and Air Force Research Laboratory (AFRL) under agreement number FA8750-20-2-1004. 
The U.S. Government is authorized to reproduce and distribute reprints for Governmental purposes notwithstanding any copyright notation thereon. The views and conclusions contained herein are those of the authors and should not be interpreted as necessarily representing the official policies or endorsements, either expressed or implied, of DARPA, 
AFRL or the U.S. Government. Address all correspondence to Edward J. Delp, \texttt{ace@ecn.purdue.edu}.}}
\address{
\parbox{\linewidth}{%
\centering\small
\textsuperscript{\symsword}Video and Image Processing Lab (VIPER), School of Electrical and Computer Engineering,\\
Purdue University, West Lafayette, Indiana, USA}\\
\parbox{\linewidth}{%
\vspace*{1ex}
\centering\small
\textsuperscript{\symshield}Dipartimento di Elettronica, Informazione e Bioingegneria, Politecnico di Milano, Milano, Italy}
}

\pagestyle{plain}

\begin{document}
\KOMAoptions{fontsize=10pt}

\maketitle

\begin{abstract}
Audio signals are often stored and transmitted in compressed formats.
Among the many available audio compression schemes, MPEG-1 Audio Layer III (MP3) is very popular and widely used.
Since MP3 is lossy it leaves characteristic traces in the compressed audio which can 
be used forensically to expose the past history of an audio file.
In this paper, we consider the scenario of audio signal manipulation done by temporal splicing of compressed and uncompressed audio signals.
We propose a method to find the temporal location of the splices based on transformer networks. 
Our method identifies which temporal portions of a audio signal have undergone single or multiple compression at the temporal frame level, which is the smallest temporal unit of MP3 compression.
We tested our method on a dataset of 486,743 MP3 audio clips.
Our method achieved higher performance and demonstrated robustness with respect to different MP3 data when compared with existing methods.
\end{abstract}

\begin{textblock}{12.5}(1.5,0.2)
\bgroup
\color{gray}
\fontsize{8}{8}\selectfont
\sffamily
\vspace*{1cm}
\noindent\textcopyright~2022 IEEE. Personal use of this material is permitted. Permission from IEEE must be obtained for all other uses, in any current or future media, including reprinting/republishing this material for advertising or promotional purposes, creating new collective works, for resale or redistribution to servers or lists, or reuse of any copyrighted component of this work in other works.
\par
\egroup
\end{textblock}


\section{Introduction}\label{sec:intro}


The advance in machine learning techniques makes generating high-quality speech or music possible \autocite{prenger2019waveglow, ren2020fastspeech, dhariwal2020jukebox}.
Almost everyone has the ability to tamper with audio signals due to the availability 
of audio editing techniques.
It is easy to synthesize/manipulate  ``fake'' speech signals that 
resembles the style and voice of a given person, and to 
concatenate real and synthetic signals to create new forged speech signals.
This poses significant threats to individuals, organizations, society and national security.

The detection of synthesize/manipulated speech is usually challenging because of the flexibility of human voice, the presence of acoustic noise or reverberation, and the complexity of audio synthesis models \autocite{borrelli2021synthetic}.
However, audio signals are mostly saved and shared using lossy compression techniques.
Lossy compression leaves distinct artifacts in the compressed audio which can be used for forensic analysis \autocite{yang2009defeating,liu2010detection}.
In this paper we examine the forensic problem of detecting if an audio signal has been spliced into another audio signal by detecting locations in the signal that have been multiply compressed.
The spliced audio signal may be real or synthetic.

Introduced in 1993, MPEG-1 Audio Layer III (MP3) \autocite{iso-13818-3, musmann2006genesis} has been one of the most popular digital audio compression methods.
It changed our ways of listening to music, podcasts, and many other types of audio content. 
Despite having inferior compression efficiency compared to \gls{aac} \autocite{iso-13818-7}, MP3 is still widely used due to compatibility with many existing applications and lower computational complexity \autocite{brandenburg1999mp3}.

Most of the existing MP3 audio forensics methods focus on double MP3 compression detection.
These methods predict whether an entire MP3 file is compressed more than once.
In \autocite{yang2009defeating,liu2010detection, yang2010detecting, qiao2013improved}, the authors used different statistical and feature design techniques on the \gls{mdct} coefficients for double compression detection.
\textcite{ma2014detecting} used the statistics of scalefactors for detecting doubly compressed MP3 with the same data rate.
In \autocite{ma2014huffman}, the authors used the Huffman table indices for double compression detection.
\textcite{yan2018compression} addressed the problem of double and triple compression detection using features extracted from scalefactors and Huffman table indices.
In \autocite{bianchi2014detection}, the authors addressed the MP3 multiple compression localization problem for the first time.
Essentially, their proposed method detects double compression using the histogram of \gls{mdct} coefficients.
Localization was achieved by using small sliding detection windows.
This technique allows one to extend any detection method to a localization method.
Finally, more modern methods make use of \glspl{dnn}.
This is the case of \textcite{luo2020compression} that proposed to use stacked autoencoder networks to detect multiply compressed audio signals.

In this paper, we present an MP3 multiple compression localization technique based on deep transformer neural networks \autocite{vaswani2017attention}.
Given an MP3 signal, our proposed method can distinguish between single compressed temporal segments and multiple compressed temporal segments thereby allowing us to temporally localize where the audio signal may
have been spliced.
Our proposed method can also be used for synthesized audio detection.


\section{Background}\label{sec:background}
We first introduce a basic overview of MP3 compression. More details of MP3 compression can be obtained in \autocite{iso-13818-3,brandenburg1999mp3, musmann2006genesis}.
Then we provide an overview of transformer neural networks.

\subsection{MP3 Compression}
\begin{figure}[htpb]
\centering
\includegraphics[width=\linewidth]{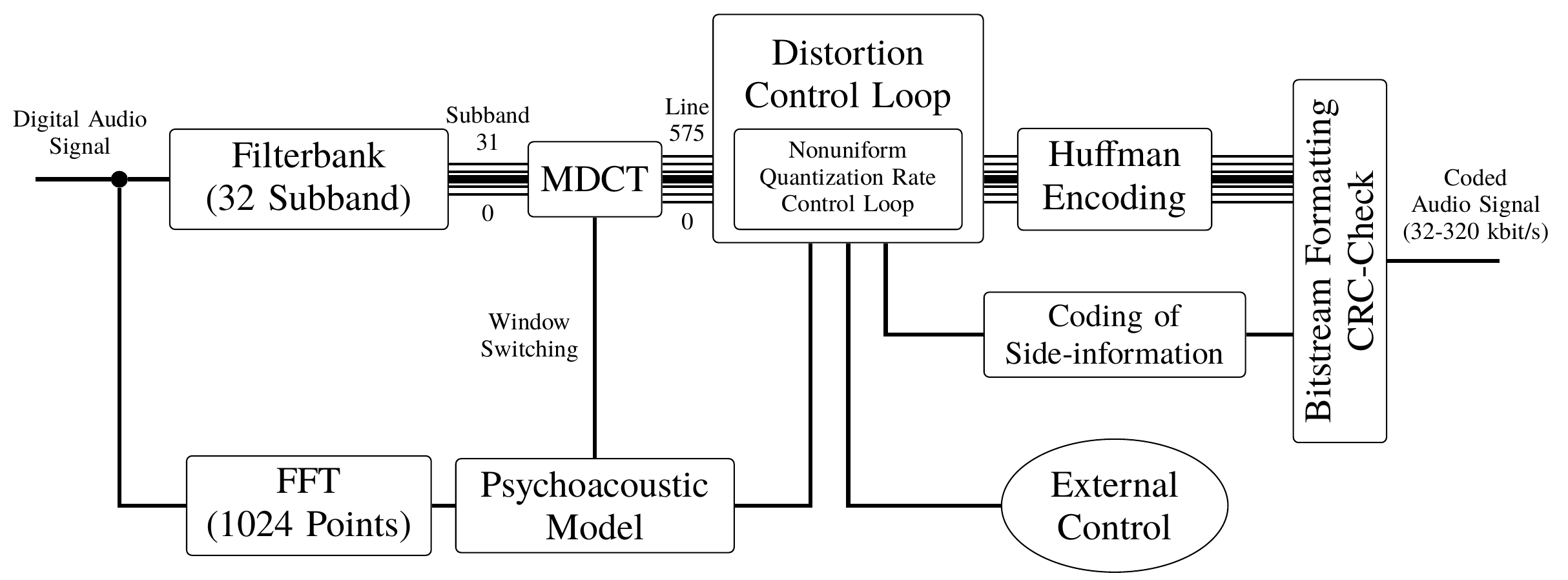}
\caption{The block diagram of an MP3 encoder.}
\label{fig:mp3-encoder}
\end{figure}

The block diagram  of a typical MP3 encoder is shown in \cref{fig:mp3-encoder}.
To compress a digital audio signal using MP3, the signal is first split into fixed time length sample windows
known as frames, where each frame contains 1152 temporal samples \autocite{iso-13818-3, musmann2006genesis}.
MP3 files are made up of a series of such frames.


The input is first processed through a perceptual model that drives the selection of coding parameters (lower branch of \cref{fig:mp3-encoder}).
During this step, the audio samples are transformed into frequency domain using \gls{fft}.
The \gls{fft} magnitude is passed to the psychoacoustic model, which exploits the characteristics of \gls{hhs} to balance between the sound quality and the data rate of the compressed signal \autocite{jayant1993signal}.
After this perceptual audio analysis, the lossy coding step is next (upper branch of \cref{fig:mp3-encoder}).
The temporal samples are filtered into 32 equally spaced frequency sub-bands using a polyphase filterbank.
Each sub-band is windowed according to the psychoacoustic model to reduce artifacts and then transformed through \gls{mdct}, which leads to 18 coefficients. 
In total, there are $32\times 18=576$ \gls{mdct} coefficients. 

After the \gls{mdct}, the resulting coefficients are quantized in the distortion control loop exploiting the psychoacoustic analysis.
The \gls{hhs} has approximately 24 critical bands based on a model of the \gls{hhs} \autocite{jayant1993signal}.
During quantization, the 32 sub-bands are grouped into scalefactor bands, which approximates the critical bands of \gls{hhs}.
The MDCT coefficients in each scalefactor band is multiplied by a scalefactor before quantization. The quantization step size increases as the frequencies become greater, because the \gls{hhs} is less sensitive to higher frequency. The scalefactors and quantization step sizes work together to satisfy both audio quality and data rate constraints. 
After quantization, the MDCT coefficients are binary encoded using one of the 32 predefined Huffman tables.
The binary coded coefficients, the encoding parameters (side-information) such as scalefactors, Huffman table indices, quantization step sizes are inserted in the data stream to form the compressed audio signal.

\subsection{Transformer Neural Networks}\label{sec:tf-nn}

\smesetsym{bm}{q,k,v,Z,A,U}
\smenewenv{align}
\makeatmath
\newcommand{\dmodel}{d_{\mathrm{model}}}
\newcommand{\atdh}{d_h}

Transformer networks have shown excellent performance in a variety of tasks such as language modeling \autocite{brown2020language}, image classification \autocite{dosovitskiy2020image}, object detection  \autocite{carion2020end}, protein structure prediction \autocite{jumper2021highly}. 

Let the input to the transformer network be @Z \in \mathbb{R}^{N \times \dmodel}@, which contains @N@ elements, and each element is a vector of @\dmodel@ dimensions.
Transformer networks use the \gls{sa} mechanism \autocite{vaswani2017attention, dosovitskiy2020image} to exploit the relationship between different elements in the input. 
Linear projection @U_{\smeraw{qkv}} \in \mathbb{R}^{\dmodel \times 3\atdh}@ is first used to generate three different projected versions of the input, i.e., @q@, @k@, and @v@:
\begin{smealign}
\left[q \mid k \mid v\right] &= ZU_{\smeraw{qkv}}.
\end{smealign}
Then, the vectors @q@ and @k@ are used to form the attention map @A \in \mathbb{R}^{N \times N}@:
\begin{smealign}
A &= \smeraw{\operatorname{softmax}} \left( \nicefrac{qk^T}{\sqrt{\atdh}} \right).
\end{smealign}
Finally, the \gls{sa} of @Z@ is the matrix multiplication of @A@ and @v@:
\begin{smealign}
\smeraw{\operatorname{SA}}(Z) = Av.
\end{smealign}
The transformer networks we use are based on \gls{msa} \autocite{vaswani2017attention}, which is an extension of \gls{sa} where @h@ different \gls{sa} values (called ``heads'') are computed in parallel. The @h@ different \gls{sa} values are combined to the result of \gls{msa} using the matrix @U_{\smeraw{msa}} \in \mathbb{R}^{\dmodel \times \dmodel}@:
\begin{smealign}
\smeraw{\operatorname{MSA}}(Z) = \left[
\smeraw{\operatorname{SA}}_1(Z) \mid
\cdots
\mid \smeraw{\operatorname{SA}}_h (Z)
\right] U_{\smeraw{msa}}.
\end{smealign}

One must guarantee $h$ divides $\dmodel$ and set @\atdh = \nicefrac{\dmodel}{h}@ so that @\smeraw{\operatorname{MSA}}(Z)@ and @Z@ have the same dimensionality. 
More details about transformers can be found in \autocite{vaswani2017attention, dosovitskiy2020image}.

%

\makeatletter

\section{Multiple MP3 Compression Localization}\label{sec:method}
In this section we first introduce the temporal splicing detection and localization problem. 
We then present our proposed solution.

\subsection{Problem Formulation}
In MP3 compression, each audio signal is composed by a series of nonoverlapping fixed temporal length segments known as frames.
Let $\x$ be the audio signal 
$\x = \{x_1, x_2, \ldots, x_L\}$,
where $x_l$ is the $l$-th frame of the signal, and $L$ is the total number of frames, which depends on the signal length.
We can associate to the audio file a sequence of labels $\y$ defined as
$\y = \{y_1, y_2, \ldots, y_L\}$,
where $y_l$ is the $l$-th binary valued label indicating whether the $l$-th frame ($x_l$) has been compressed once (i.e., $y_l=0$) or more than one time (i.e., $y_l=1$).

During the creation of a spliced MP3 audio file, it is likely that frames from different audio signals are concatenated in time and compressed using MP3.
Some of the audio signals can be uncompressed (e.g., pristine or generated from a synthetic speech),  
while others may have been compressed and then decompressed.
The final spliced signal will likely contain both single and multiple compressed frames.

Our goal is to find $\yhat$, which is an estimate of the sequence of labels $\y$ associated with the audio file $\x$ by examining the MP3 compressed data for $\x$ on a frame by frame basis.
In doing so, we are able to detect if an audio file has undergone splicing, and localize which frames have been compressed more than one time.

\subsection{Proposed Method}

\begin{figure*}[t]
\centering
\includegraphics[width=0.81\linewidth]{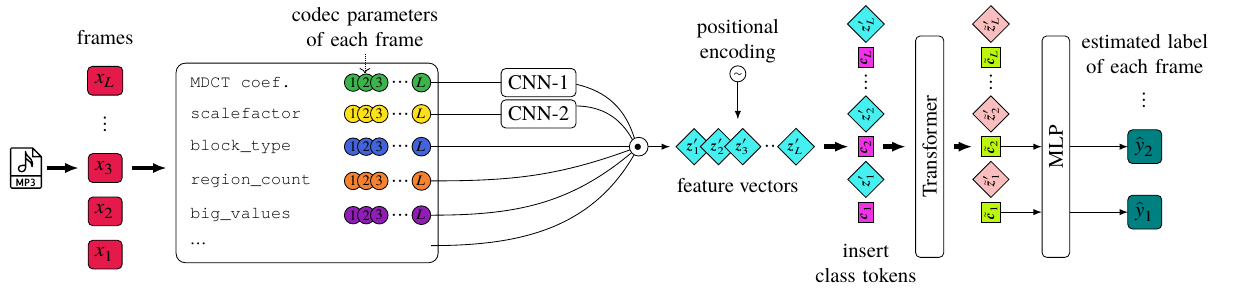}
\input{./figures/architecture/architecture_export.tex}
\caption{The block diagram of the proposed method, which analyzes $L$ frames at a time. Each circular node (\protect\vecnode{}{$l$}) represents a MP3 codec parameter vector whose corresponding frame is shown by the number inside. Each diamond node (\protect\diamondnode{}{$\z'_l$}) represents a vector associated with the feature vector $\z'_l$. Each rectangular node (\protect\clsnode{}{$\c_l$}) represents a vector associated with the class token $\c_l$. 
Different groups of vectors are shown in different colors.
\vspace*{-2em}
} 
\label{fig:method-design}
\end{figure*}

Our proposed method is shown in \cref{fig:method-design}.
We examine the MP3 data corresponding to the individual frames of the input 
signal $\x=\{x_1, \, x_2, \, ..., x_L\}$.
We examine $L$ frames at a time to decide if the audio signal has been 
multiply compressed and localize the splicing.
A MP3 compressed frame consists of two groups of samples known as granules \autocite{iso-13818-7, musmann2006genesis}.
Each granule contains 576 temporal samples from the respective two stereo channels. 
Our proposed method uses the data produced by the MP3 codec for a frame shown in \cref{tbl:mp3-codec-fields}
from the first channel of the first granule. 
This data consists of MDCT coefficients, quatization step sizes, scalefactors, Huffman table indices, and sub-band window selection information.
The MP3 codec data we choose contain a complete set of parameters required to decode the channel.
We will use this information to decide if an audio signal has been multiple compressed and localize the muliple compressed frames.

The \texttt{mdct\_coef} and \texttt{scalefactor} fields are used as inputs to two separate \glspl{cnn} (i.e., CNN-1 and CNN-2) to find more features (see \cref{fig:method-design}). 
We use CNN architectures that are similar to the well known VGG CNN \autocite{liu2015very}. 
The depth and number of filters are changed based on the size of the \texttt{mdct\_coef} and \texttt{scalefactor} fields.
By denoting the convolutional layer as Conv<receptive field size>--<number of filters> and the fully connected 
layers as FC--<number of neurons>, the architecture and parameters of each \gls{cnn} are as follows:
\begin{itemize}
\item CNN-1: {\small Conv3-32, Conv3-32, Maxpool, Conv3-64, Conv3-64, Maxpool, Conv3-128, Conv3-128, Maxpool, FC-233, Dropout, FC-233, Dropout}.
\item CNN-2: {\small Conv3-16, Conv3-16, Maxpool, Conv3-32, Conv3-32, Maxpool, Conv3-64, Conv3-64, Maxpool, FC-49, Dropout, FC-49, Dropout}.
\end{itemize}


\begin{table}[htpb]
\centering
\caption{MP3 codec information fields used in our method. Details about each field can be obtained from \protect\autocite{raissi02thetheory,sripada2006mp3}.}
\label{tbl:mp3-codec-fields}
\resizebox{\linewidth}{!}{
\renewcommand{\arraystretch}{0.8}
\begin{tabular}{>{\fontsize{7}{5}\selectfont}m{0.5\linewidth}l}
\toprule
\multicolumn{1}{c}{Fields} & \multicolumn{1}{c}{Description} \\ \midrule
\texttt{part\_23\_length} & Size of coded binary data \\ \midrule
\texttt{scalefactor}, \texttt{scalefac\_compress}, \par\texttt{scalefac\_scale}, \texttt{preflag}  & Scalefactor value info \\ \midrule
\texttt{global\_gain}, \texttt{subblock\_gain}, 
\par\texttt{big\_values}, \texttt{region\_count}  & Quantization step sizes \\ \midrule
\texttt{table\_select}, \texttt{count1\_table} & Huffman table selection info \\ \midrule
\texttt{block\_type}, \texttt{mixed\_block\_flag} & Sub-band window selection info\\ \midrule
\texttt{mdct\_coef}  &  Decoded MDCT coefficients \\ \bottomrule
\end{tabular}
}

\end{table} 


For the $l$-th frame, we concatenate the outputs of CNN-1 and CNN-2 as well as the values of the remaining fields into a feature vector $\z_{l} \in \mathbb{R}^{\dmodel}$, which is used as the input to the transformers. The sizes of the CNN outputs are selected so that the dimensionality of $\z_{l}$'s satisfies $\dmodel = 300$. 
The self attention mechanism does not use the order information of the elements in the input sequence.
To allow the transformer to make use of the temporal order of frames, the frames' corresponding feature vectors are added with positional encoding \autocite{vaswani2017attention}. 
The positional encoding is a series of $L$ distinct fixed-valued vectors $\{\p_1, \p_2, \ldots, \p_L\}$, where $\p_l \in \mathbb{R}^{\dmodel}$. 
After adding positional encoding, the $l$-th feature vector is $\z'_{l} = \z_l + \p_l$. The transformer will likely be able to find the position of $\z'_{l}$ being $l$ based on the $\p_l$ component of $\z'_{l}$.


\newcommand{\clstok}{\texttt{CLS}}

We use a similar approach as described in \autocite{dosovitskiy2020image} to binary classify the feature vectors corresponding for each MP3 frame as being ``single compressed'' or ``multiple compressed''.
The network has $L$ special vectors, known as ``class tokens'', which are denoted by $\{\c_1, \c_2, \ldots, \c_L\}$, where $\c_l \in \mathbb{R}^{\dmodel}$.
The input to the transformer $\Z$ is the feature vector $\z_{l}$'s interleaved with the class tokens, which can be written as 
\begin{align}
\Z = \left[\c_1 \mid \z'_1 \mid  \c_2 \mid  \z'_2 \mid  \ldots  \mid \c_L \mid \z'_L \right] \in \mathbb{R}^{2L \times \dmodel}. 
\end{align}
Let $\tfop(\cdot)$ be the function corresponding to the transformer network. 
After the transformer operations, the output can be written as
\begin{align}
\tfop(\Z) = \left[\tilde{\c}_1 \mid \tilde{\z}'_1    \mid \tilde{\c}_2 \mid  \tilde{\z}'_2 \mid \ldots  \mid \tilde{\c}_L \mid \tilde{\z}'_L \right] \in \mathbb{R}^{2L \times \dmodel}.  
\end{align}
To find the estimated labels for the $l$-th frame, we use a \gls{mlp} network to classify $\tilde{\c}_l$.
During training, the gradients with respect to $\c_{l}$'s are computed and used to update them using gradient descent \autocite{ruder2016overview}. 
That is, we train each class token to collect information necessary to 
determine the label for its corresponding time step. 
Finally, the label of each frame can be determined by only examining the class tokens.



{
After preliminary experiments, we used 8 transformer layers with the number of heads $h=15$ in the Multihead Self Attention (see Section~\ref{sec:tf-nn}), which yielded the best performance.
}

The \gls{mlp} network we use is made up of one layer of 800 neurons and a final output layer with 2 neurons using softmax activation.
We use the GELU \autocite{hendrycks2016gaussian} nonlinear activation function in the CNNs and the transformer network.

After training, for a given MP3 audio frame sequence $\{x_1, \allowbreak x_2, \allowbreak \ldots, \allowbreak x_L\}$, our method generates a binary decision sequence $\{\hat{y}_1, \hat{y}_2, \ldots, \hat{y}_L\}$ describing whether each frame is multiply compressed or not.
This enables one to localize the questionable frames in an MP3 file and determine which part of the MP3 frames may have been spliced.


\section{Experiments and Results}\label{sec:experiments}
In this section we describe our experiments and present results comparing our method to other methods for splicing localization using traces of multiple MP3 compression.


\subsection{Dataset}

The data for our experiments contain only uncompressed WAV audio files. 
We used three publicly available datasets consisting of speech and different music genres: LJSpeech \autocite{ito2017ljspeech}, GTZAN \autocite{tzanetakis2002musical}, and MAESTRO \autocite{hawthorne2018enabling}.
We then compressed the WAV files using MP3.
The MP3 sampling rate we selected is 44.1 kHz, and the length of
each frame is $\nicefrac{1152}{44100} \approx 26.12\mathrm{ms}$.
In our experiments, our method examined $L=20$ frames at a time.

The MP3 compression data rates and data rate types used in our experiments are shown in \cref{tbl:compression-type}.
We used Constant Bit Rate (CBR) and Variable Bit Rate (VBR) compression.
The audio signals were compressed using FFmpeg\footnote{\url{https://www.ffmpeg.org/}} v3.4.8 and LAME\footnote{\url{https://lame.sourceforge.io/}} v3.100. 
We excluded low-quality MP3 compression configurations that make multiple-compression detection unreliable \autocite{yan2018compression}. 
We also excluded ultra-high-quality MP3 compression that is not recommended by FFmpeg \autocite{ffmpegmp3}.

\begin{table}[htb]
\centering
\caption{MP3 compression types used in our experiments. The data rate of VBR compression decreases as quality index increases. More details can be obtained from \autocite{ffmpegmp3}.}
\label{tbl:compression-type}
\resizebox{0.7\columnwidth}{!}{
\begin{tabular}{c>{\centering\arraybackslash}m{0.5\linewidth}}
\toprule
Compression type & Bit rate/Quality\\ \midrule
\makecell{Constant Bit Rate (CBR)} &  64kbps, 96kbps, 128kbps,\par 160kbps, 192kbps, 256kbps \\ \midrule
\makecell{Variable Bit Rate  (VBR)} & 1, 2, 3, 4, 5, 6 \\ \bottomrule
\end{tabular}
}
\end{table}

\begin{figure}[htb]
\centering
\resizebox{\linewidth}{!}{
\includegraphics{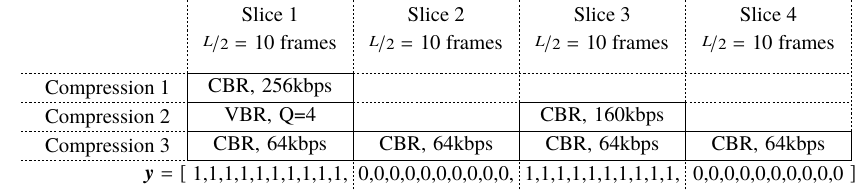}
}
\caption{An illustration of our dataset generation method on a segment of 40 frames (4 slices). The corresponding label for this segment $\y$ is shown at the bottom.}
\label{fig:dataset-gen}
\end{figure}

Our dataset generation scheme is described as follows. 
We split each uncompressed WAV audio file into smaller segments of 80--320 frames randomly.
Each segment is further temporally subdivided into ``slices'' of $\nicefrac{L}{2}=10$ frames where $L$ is the number of frames our method examines at a time.
Now, each segment will contain 8--32 slices.
In each segment, the odd slices will be multiple compressed for 2--3 times. 
The even slices will be single compressed. 
The compression parameters are chosen from \cref{tbl:compression-type} at random.
In \cref{fig:dataset-gen} we illustrate the dataset generation method on a segment containing 4 slices.
If a slice needs to be compressed three times, then compression 1--3 will be used.
If a slice needs to be compressed two times, then compression 2--3 will be used.
Otherwise, only compression 3 will be used.
For a given slice, after compression 1 the MP3 file is decompressed back to the time domain, which is used as 
the input to compression 2.
At compression 3, we gather the decompressed or pristine time domain samples of all slices in a segment and then compress them as one MP3 file.
Therefore, the parameters for compression 3 will be the same for all slices in a given segment.
This completes the construction of our ground-truthed experimental data set.

\subsection{Preparing an MP3 file For Training}
\label{sec:dataset-train}

Recall our method operates in the ``MP3 domain'' by examining the MP3 codec fields shown in 
Table~\ref{tbl:mp3-codec-fields}.
To train our method,
we used a sliding window of length $L$ with offset step size 8 to extract the MP3 codec fields from the MP3 compressed frames as shown in
Table~\ref{tbl:mp3-codec-fields}.

We generated 486,743 MP3 frame sequences of length $L$ from our experimental dataset.
{
We used 54\% of the frame sequences in the training set; 13\% of the frame sequences in the validation set, and 33\% of the frame sequences in the testing set.}
{
The partition was done in a way that all frame sequences generated from one segment can only be used in one of the three sets.
}



Selecting the MP3 frame sequence length $L$ for analysis is a trade-off between the accuracy
 of the estimation of $\yhat$ and the training difficulty of the network. 
Using a larger $L$ may improve the performance, but it may also significantly increase training time and the need of hyperparameter tuning. 
In our experiments, we used $L=20$ corresponding to
an audio signal length of approximately 522.4ms.
This may be small compared to the typical length of most audio signals. 
In training we used a sliding window stride of 8, which does not align with the slice boundaries. 
In each sliding window, the index of the first multiple compressed frame and the number of multiple compressed frame are different.
This forces the trained network to predict the labels correctly given an arbitrary portion taken from an audio signal. 
Therefore, our method is able to examine  longer audio files even if $L$ is relatively small.

\subsection{Hyperparameters and Training}

The entire network including the CNNs and the transformers are trained end-to-end. 
We trained the network using the Adam optimizer \autocite{kingma2014adam} with an initial learning rate of $10^{-4}$  and 
a dropout rate of 0.2
until the validation accuracy no longer increased for 20 epochs.

\subsection{Results}

We compared the performance of our method to \autocite{liu2010detection, yang2010detecting, yan2018compression}, which are introduced in Section \ref{sec:intro}.
Since they are all detection methods, we used the predictions from these methods on short frame sequences of length
$L'=4$ to approximate localization. 
Choosing the length $L'$ for localization approximation is a trade-off between classification accuracy and granularity of localization. 
{
Larger $L'$ improves the label estimation of each frame sequence, while smaller $L'$ improves the localization accuracy.
}
Selecting $L'=4$ is a reasonable balance between these two factors \autocite{bianchi2014detection}.

In \cref{tbl:perf-metric-comparision}, we compare the performance of our method to that of the three previous methods. 
We used three different metrics: Jaccard score \autocite{levandowsky1971distance}, $F_1$-score \autocite{powers2020evaluation} and balanced accuracy score \autocite{brodersen2010balanced}.
Let $\ifunc(\y)$ be the function that returns the set of indices for the one-entries in $\y$. 
The Jaccard score can be computed by $\nicefrac{|\ifunc(\y) \cap \ifunc(\yhat)|}{|\ifunc(\y) \cup \ifunc(\yhat)|}$, which compares the similarity between the predicted multiple compressed region and the ground truth multiple compressed region \autocite{levandowsky1971distance}. 
The $F_1$-score is the harmonic mean of the precision and recall, which considers both factors at the same time \autocite{powers2020evaluation}. 
The balanced accuracy score is the traditional accuracy score with class-balanced sample weights \autocite{brodersen2010balanced}.
Our approach achieved the highest score on all three metrics. 
{
We also tested our method on a separately generated dataset containing 53,541 MP3 frame sequences with variable slice length ranging from 10 to 80 frames.
For each slice, the slice length, compression types and the number of compressions are chosen at random.
Our method achieved 81.64 balanced accuracy on this dataset, which is close to the result in \cref{tbl:perf-metric-comparision}.
This shows our method did not learn strong dataset bias.
}

In \cref{tbl:recall-last-comp-type-comparision}, we show the recall \autocite{powers2020evaluation} of each method on multiply compressed MP3 frames against selected last MP3 compression types. 
It can be seen that the detection accuracy decreases as the MP3 compression quality declines. 
Different methods reacted to CBR and VBR compression in contrasting manners. 
For \autocite{liu2010detection} and our method, the performance was similar.
For \autocite{yang2010detecting}, the score of VBR frames was significantly lower than those of CBR; the behavior of \autocite{yan2018compression} was the opposite. 

In table \ref{tbl:recall-num-compression}, we show the recall of each method compared to the number of MP3 compression applied to a frame. The recall of our method is more consistent across different repetition of MP3 compression. 
For all methods, the recall for double compression is close to that of triple compression. 

Overall, our approach demonstrated high localization performance and robustness across many types of MP3 compression.

\begin{table}[htb!]
\centering
\caption{Performance metrics comparison.}
\label{tbl:perf-metric-comparision}
\resizebox{0.6\columnwidth}{!}{
\begin{tabular}{lccc}
\toprule
\multicolumn{1}{c}{Method} & \makecell{Jaccard\\Score}  & $F_1$-score & \makecell{Balanced\\ Accuracy}\\ \midrule
\textcite{yan2018compression} & 13.53 & 18.71 & 53.35 \\
\textcite{yang2010detecting} & 30.73 & 40.95 & 55.28 \\
\textcite{liu2010detection} & 48.18 & 58.91 & 68.72 \\
Our Approach & 80.50 & 84.43 & 84.49 \\ \bottomrule
\end{tabular}
}
\vspace*{2ex}
\end{table}

\begin{table}[htb!]
\centering
\caption{The recall of multiple compression localization of each method against selected last MP3 compression types. CBR compression is denoted by C<bit rate>; VBR compression is denoted by V<quality index>.}
\label{tbl:recall-last-comp-type-comparision}
\resizebox{0.8\columnwidth}{!}{
\renewcommand{\tabcolsep}{1ex}
\begin{tabular}{l*{8}{c}}
\toprule
& \multicolumn{8}{c}{Last MP3 Comression Type} \\ \cmidrule{2-9}
\multicolumn{1}{c}{Method} & C64 & C128 & C160 & C192 & V1 & V2 & V4 & V6\\ \midrule
\textcite{yan2018compression} & 17.30 & 6.86 & 6.80 & 4.87 & 22.80 & 23.23 & 22.59 & 17.55 \\
\textcite{yang2010detecting} & 19.27 & 70.75 & 71.71 & 66.21 & 45.58 & 39.05 & 27.12 & 22.06 \\
\textcite{liu2010detection} &  58.45 & 56.48 & 54.47 & 55.01 & 55.15 & 56.41 & 57.47 & 67.93 \\
Our Approach & 73.09 & 88.06 & 89.65 & 90.84 & 93.31 & 91.21 & 83.52 & 65.51\\ \bottomrule
\end{tabular}
}
\vspace*{2ex}
\end{table}

\begin{table}[htb!]
\centering
\caption{The recall of each method against the number of MP3 compression.}
\resizebox{0.65\columnwidth}{!}{
\begin{tabular}{lcccc}
\toprule
 & \multicolumn{3}{c}{Num. MP3 Compression} & \\ \cmidrule{2-4}
\multicolumn{1}{c}{Method} & Single & Double & Triple & Overall\\ \midrule
\textcite{yang2010detecting} & 67.28 & 43.51 & 43.02 & 43.27 \\
\textcite{yan2018compression} & 91.71 & 14.86 & 15.10 & 14.98 \\
\textcite{liu2010detection} & 79.36 & 57.27 & 58.85 & 58.06 \\
Our Approach & 84.61 & 83.76 & 84.92 & 84.34\\ \bottomrule
\end{tabular}
}
\label{tbl:recall-num-compression}
\vspace*{2ex}
\end{table}

\section{Conclusions}\label{sec:conclusions}
We proposed a multiple MP3 compression temporal localization method based on transformer neural networks that uses MP3 compressed data.
Our proposed method localizes multiple compression at the frame level. 
The experiment results showed that our method had the best performance compared to other approaches and 
was robust against many MP3 compression settings. 
In the future, we will examine extending this approach to other compression methods such as AAC.
We will also investigate the use of stereo channels as well as the second granule in MP3 compressed frames.
We will generalize the concept of multiple compression detection to compression history detection. 
That is, to find the number of compressions and the types of compression used.
Knowing the compression history can greatly enhance the interpretability for audio forensics.


\FloatBarrier





\clearpage

\section{References}
{

\AtNextBibliography{
  \fontsize{9}{10}\selectfont
}
\setlength\bibitemsep{0.2ex}
\printbibliography[heading=none]
}

\end{document}